\documentclass{article}

\usepackage{bioarxiv}
\usepackage[utf8]{inputenc} % allow utf-8 input
\usepackage[T1]{fontenc}    % use 8-bit T1 fonts
\usepackage{hyperref}       % hyperlinks
\usepackage{url}            % simple URL typesetting
\usepackage{booktabs}       % professional-quality tables
\usepackage{amsfonts}       % blackboard math symbols
\usepackage{nicefrac}       % compact symbols for 1/2, etc.
\usepackage{microtype}      % microtypography
\usepackage{graphicx}
\usepackage{wrapfig}
\usepackage{amsmath}
\usepackage{multicol}        % used for the two-column index
\usepackage[bottom]{footmisc}% places footnotes at page bottom
\usepackage{newtxtext}   
\usepackage{newtxmath}
\usepackage{makecell}
\usepackage{adjustbox}

\usepackage{subfig}
\usepackage{tabularx}
\usepackage{longtable}
\usepackage{multirow}
\usepackage{tabularx}
\usepackage{lscape}
\usepackage{setspace}

\title{Ten Simple Rules for Success with HPC, i.e. Responsibly BASHing that Linux Cluster}

\author{
  Jamie J. Alnasir\\
  Department of Computing\\
  Imperial College\\
  London, UK \\
  \texttt{j.alnasir@imperial.ac.uk} \\
}

\begin{document}

\maketitle

\keywords{HPC \and training \and BASH \and workflow languages \and best-practices}

\section*{Introduction}

High-performance computing (HPC) clusters are widely used \textit{in-house} at scientific and academic research institutions \cite{urlEPSRC-HPC}. For some users, the transition from running their analyses on a single workstation to running them on a complex, multi-tenanted cluster, usually employing some degree of parallelism, can be challenging, if not bewildering, especially for users whose role is not predominantly computational in nature. On the other hand, there are more experienced users, who can benefit from pointers on how to get the best from their use of HPC.

This Ten Simple Rules guide is aimed at helping you identify ways to improve your utilisation of HPC, avoiding common pitfalls that can negatively impact other users and will also help ease the load (pun intended) on your HPC sysadmin. It is intended to provide technical advice common to the use of HPC platforms such as LSF, Slurm, PBS/Torque, SGE, LoadLeveler and YARN, the scheduler used with Hadoop/Spark platform.

\section*{Rule 1: Do the HPC induction! Read the manual!}

Our first rule starts with a fitting anecdote to a first encounter with HPC in my first year of my Ph.D. During lunch with colleagues, and after a lull in the conversation, Bob --- the IT support lead for our department --- exclaimed: "I just received 14,000 emails thanks to a HPC user when I arrived at my desk this morning" which of course got the conversation going again with some intrigue and amusement.  He was referring to the automatic emails sent out by the department's HPC cluster's job-scheduler, IBM LSF (Load Sharing Facility) --- it was configured to automatically email the job-submitting user an email report on job completion \cite{urlLSFemail}.  Failing that, e.g. when the user's inbox is already full, it is set-up to email the report to the designated sysadmin. It transpired that a certain user --- yes, you've guessed correctly, it was me --- submitted a very large number of jobs (one for each entry in the Protein Data Bank), but had not overridden the default setting to email the job report back to the user.

Whilst the automatic emailing of job-status reports as a default is not a common configuration, it's not unheard of. It wasn't immediately obvious to me that I needed to specify an output filename using the -o option to \textit{bsub}, in order to suppress the job report email and write the output to a file instead. Hence, the first rule here is: \textit{1. Do the HPC induction! Read the manual!}

The cluster I had used belonged to a small research group. It had only a few users and there wasn't any induction training or manual. However, most institutions provide an HPC induction and training material, often also available via an institutional intranet, because these HPC clusters are typically multi-tenanted systems and improper use can negatively impact other users. It makes sense to make sure you do this before you commence.

\section*{Rule 2: Brush up on your Linux commands and BASH!}

BASH (Bourne-Again SHell) is a \textit{shell} --- a  command-line interpreter language --- whose name is a pun on Stephen Bourne, the developer of the UNIX shell, sh which BASH descends directly from \cite{ramey1994bash}. A shell provides access to the Operating System's services. Areas to which particular focus should be given include, for example:

\begin{itemize}
    \item \textbf{Streaming, piping, redirection}: these facilitate the control of input to and output from processing.
    \item \textbf{Environment variables}: for setting and sharing of parameters between software as well as the OS.
    \item \textbf{Compression and archiving}: usually necessary prior to network transfer, useful for collating large numbers of files in more manageable single archives, and for optimising storage.
\end{itemize}

\section*{Rule 3: Don't simply run your commands on a login node!}

\begin{figure}[ht]
    \centering
    \includegraphics[width=0.72\linewidth]{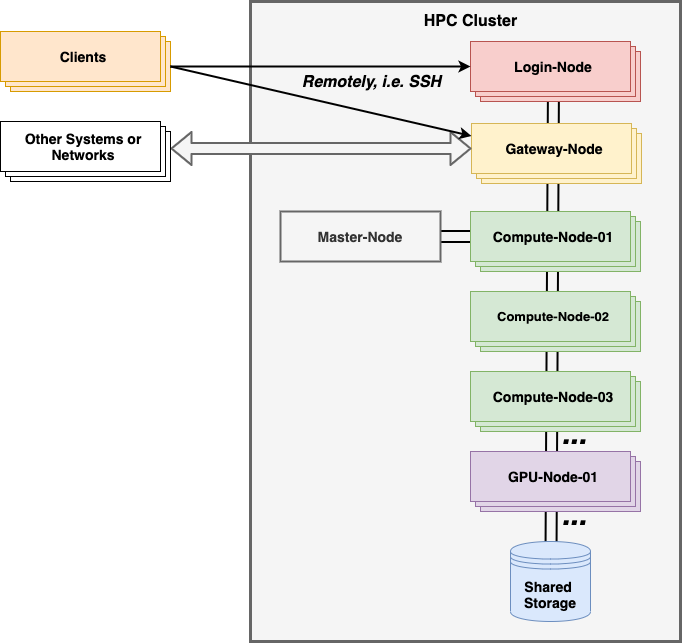}
    \caption{Simplified organisation of a typical HPC system. Users log in remotely via SSH to a \textit{login} node from a client machine. Jobs are scheduled by job-scheduler software running on the \textit{master} node and run on \textit{compute} nodes and \textit{GPU} nodes. \textit{Gateway} node(s) facilitate the transfer of large files into and out of the HPC system.}
    \label{fig:HPC}
\end{figure}

There are often one or more login nodes which are reserved to facilitate users logging on to the cluster via SSH and for submitting jobs to the job-scheduler, as well as performing non-intensive tasks, such as editing source-code and job submission scripts (Figure \ref{fig:HPC}). This is because the load on the cluster is balanced by the job-scheduler running on the master node. This is the reason why logging into individual nodes to run jobs directly on them is bad practice and is usually not permitted. Login nodes are typically light-weight Virtual Machines and are therefore not suitable for running intensive tasks, doing so would slow the login nodes for other users and cause delays in logging in. It is common practice for such tasks to be terminated often without prior notice.

Job-schedulers typically provide an \textit{interactive job} mode where an interactive shell session is initiated on one of the compute nodes and responds to both ad hoc commands and scripts running within the session. This can be used for slightly more intensive tasks such as e.g. compiling code for HPC, recursively finding files.

\section*{Rule 4: Use a workflow language for your analysis pipeline!}

Although the HPC batch-cluster systems commonly used in research institutes and universities can be used interactively --- i.e. through an \textit{interactive session} --- most of the time you will implement scripts to perform certain tasks on some input data, such as when developing an analysis pipeline. A variety of different languages can be used for this, e.g. Python, Perl, or a shell scripting language, such as BASH.

The trouble with this approach, however, i.e. writing collections of scripts \textit{de novo} to do this, is that such scripts will invariably be platform specific, will need to handle parallelism (particularly managing child-jobs and their dependencies), error conditions and, in order to be reproducible, will have to be standardised in some way. Without effectively addressing these issues, which together are often non-trivial, the result is often a script that lacks interoperability, is difficult to reproduce, and will therefore pose challenges in sharing, particularly with external collaborators. A solution for this is to use a workflow language such as WDL (Workflow Definition Language) \cite{urlWDLspec}, CWL (Common Workflow Language) \cite{amstutz2016common} or Nextflow \cite{di2017nextflow}, which will allow for platform agnosticism, standardisation and reproducibility, particularly when used with a package manager.

\section*{Rule 5: Consider a package manager!}

A software package management system automates the installation, configuration, upgrading, and removal of software on an operating system. For example, Anaconda is a ready to use distribution of datascience packages together with a package manager called Conda for any programming language (most commonly Python and R), which is frequently used with Linux and often installed on HPC clusters \cite{urlConda, yan2018hands}. An initial installation is supplied with Python, Conda and over 150 scientific packages including their library dependencies. The use of a package manager obviates the need for the user to explicitly manage packages and their configuration, which can often be complex --- automatic installation and updating of packages takes care of downloading, updating and configuring dependencies which are configured in separate \textit{Conda} environments.

Another important advantage of this approach is that it allows the developer to isolate and manage version dependencies for different projects. For example, when combined with a workflow language, such as Nextflow, the reproducibility of pipelines can be assured by pinning down specific pipeline steps to use specific versions of tools as defined by designated \textit{Conda} environments. In such a case, the packages defined in a \textit{Conda} environment can be exported into a YAML file and this referenced in the workflow step that requires the specific environment. The required packages can then be loaded, or if necessary, downloaded on-the-fly by Nextflow and cached for subsequent use, thereby allowing workflow pipelines to be shared collaboratively with the confidence the correct packages will be used for execution.

\section*{Rule 6: Understand the levels of parallelism required by your job!}

In order to leverage the increased throughput that parallelism inherent in distributed computing offers, it is necessary to understand the levels on which this is applied in executing the computational tasks in your workflow.

\begin{figure}[ht]
    \centering
    \includegraphics[width=1\linewidth]{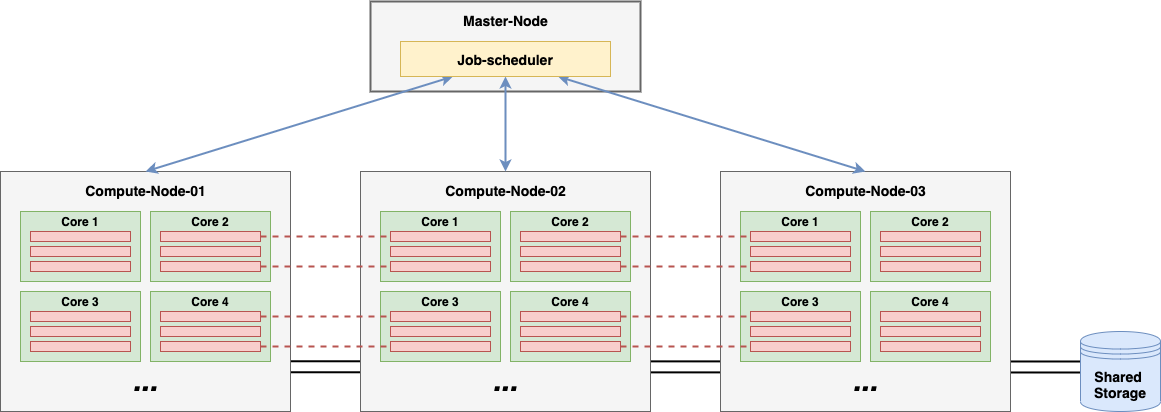}
    \caption{Parallelism in a typical HPC cluster. The job-scheduler runs on the master node and controls the highest-level of parallelism in running jobs on each node (blue arrows). For example, in genome processing, each node may perform the alignment of an individual chromosome to a reference genome and could use multiple threads per core --- 3 threads per core are depicted, for a total of 12 per chromosomal job. In MPI jobs, threads on one node can communicate with threads on other nodes through inter-process communication (depicted in red dashed lines).}
    \label{fig:HPC-parallelism}
\end{figure}

In an HPC cluster, the highest level --- that is the coarsest granularity of parallelism --- occurs at the compute node level, and the lowest level --- that is the finest granularity -- occurs at the thread level, on each CPU core (Figure \ref{fig:HPC-parallelism}). In MPI (Message Passing Interface) jobs, processes can communicate with each other across the network through message-passing which requires low-latency, high bandwidth interconnects between the compute nodes.

In the GATK (Genome Analysis Toolkit) widely used in genomics and transcriptomics analyses, earlier versions allow for the control of parallelism using the \textit{-nt}, and \textit{-nct} parameters which allow the user to specify the number of threads and number of threads per core, respectively. In GATK4, some tools are Spark-enabled, and can make use of high-throughput distributed multi-threading of the Spark platform --- in this case, parallelism is controlled at the thread level, and is set using the \textit{-XX:ParallelGCThreads} java parameter \cite{urlGATK4}. These parameters should be set according to the resources allocated for the job and will be dependent on the architecture of the HPC system in use.

\section*{Rule 7: Identifying potential bottlenecks in your job!}

The job-scheduler will produce a job report which details the execution time and resource utilisation. Ideally, the report should show high $\%$ CPU utilisation --- the time during which the CPU executed user commands, and a low $\%$ system usage --- the time during which the system had spent kernel mode executing system processes, e.g. forking, spawning processes and I/O. [Incidentally, this system time overhead is the reason care should be taken whilst submitting large numbers of very small jobs. HPC batch-schedulers usually allow such types of jobs to be submitted in an aggregated way, i.e. as an array-job so that they can optimally manage the system processes involved and scheduling].

The wall-time (or real-time) of a job is a basic metric that is reported by the HPC job-scheduler and can be observed. This is simply the time interval between the job's start and finish --- if your job is taking a long time to run on HPC, it is possible that something might be awry, and it is worth investigating further. The Linux \textit{time} command can be prepended to individual programs or commands in a batch-script and used to monitor the time taken (real-time, user-time, system-time) for that task to execute.

The types of resource bottlenecks that are likely to occur are:

    \begin{itemize}
        \item \textbf{CPU-bound}: execution is highly dependent on the speed of the CPU. Rewriting of the algorithm is usually necessary.
        \item \textbf{Disk-bound}: execution time is dependent on disk I/O, e.g. seek times. Your job may be terminated automatically to prevent \textit{Disk-thrashing} from occurring. If possible, use a faster storage media i.e. SSD. Also, caching data into memory usually helps to reduce repeated disk access.
        \item \textbf{Memory-bound}: execution time is highly dependent on the speed of memory access and is limited by the amount of available memory. Scaling up may help with increasing the available memory per process, but faster access requires faster memory.
        \item \textbf{Network-bound}: the network bandwidth is the limiting factor. Where possible, run processes on the same host so as to obviate network traffic.
    \end{itemize}

\section*{Rule 8: Take care with large-file transfers!}

It will often be necessary to transfer large files into and out of a HPC cluster, e.g. Next-generation sequencing data. Whilst HPC nodes are connected with high-speed network interconnects, the available bandwidth is finite and shared by all of the users, their jobs, the operating system and the cluster-management software. For this reason, HPC systems often have designated nodes --- typically called gateway nodes (see Figure \ref{fig:HPC}) --- specifically for file transfers into and out of the HPC cluster, with defined procedures or commands to use. Other tips for large file transfers include scheduling them to run i.e. on the appropriate gateway node or using the correct procedure, during periods of low cluster activity i.e. at night, or notifying your HPC support team who can perform or schedule them on your behalf. 

\section*{Rule 9: Be aware of any recharge model!}

Some institutions operate a recharge model whereby the computing department that runs the HPC service bills research groups or departments who use the cluster. On an HPC system, it is typically storage and disk space that are charged for. In these cases CPU usage is usually charged in CPU/hours for CPU time used --- not wall-time requested on job submission --- and storage charges are usually billed in, e.g. TB/year for exceeding a default user quota. HPC jobs are therefore often submitted with a \textit{project-code} to identify the research group or project, particularly when they are billed for use. So, it is important to know beforehand if a recharge model is in place, especially if you are going to submit a very intensive, or long duration job, or will exceed your disk quota.

\section*{Rule 10: Don't use Unicode characters in file paths!}

As much as you may want to use the correct diacritics for your name in your filenames and file paths, try to avoid this if doing so involves the use of uncommon Unicode characters. It will create problems for other users trying to access your files and will also create difficulties for the HPC support team when they come to perform non-automated operations on your files such as, for example, migrations, file transfers and routine house keeping.

{\small

\section*{Author profile}

Dr. Jamie J. Alnasir is a post-doctoral research associate at the SCALE Lab, Department of Computing, Imperial
College London. His research interests are distributed computing, high-performance computing, DNA-storage,
computational biology, next-generation sequencing and bioinformatics. At the ICR (Institute of Cancer Research) London, he worked at the Scientific Computing department helping researchers leverage HPC, training them in the use of workflow languages and consulting in scientific software engineering.

\section*{Conflicts of Interest Statement}

The author declares no conflicts of interest.

}

\bibliographystyle{unsrt}  
\bibliography{main} 

\end{document}